\documentstyle[aas2pp4]{article}
\lefthead{Kunzl et al.}
\righthead{Pair production in Crab}
\begin{document}
\title{On pair production in the Crab pulsar}
\author{Thomas Kunzl\altaffilmark{1,3}, Harald
Lesch\altaffilmark{1},Axel Jessner\altaffilmark{2} and Alexis von
Hoensbroech\altaffilmark{1,2}}
\altaffiltext{1}{Universit\"ats-Sternwarte M\"unchen, Scheinerstr.1,
D-81679 M\"unchen, Germany}
\altaffiltext{2}{Max-Planck-Institut f\"ur Radioastronomie, Auf dem
H\"ugel 69, D-53121 Bonn, Germany}
\altaffiltext{3} {Max-Planck-Institut f\"ur extraterrestrische
Physik,Giessenbachstr., D-85740 Garching, Germany}

\authoraddr{Th. Kunzl}
\authoremail{tak@usm.uni-muenchen.de}
\date{}
%\maketitle

\begin{abstract}
We consider the widespread assumption that coherent pulsar radio
emission is based on extended pair production leading to plasma
densities highly exceeding the Goldreich-Julian density.
We show as an example that the observed low frequency (160 MHz)
emission of the Crab pulsar is incompatible to the model of
extended pair production.
Our results rule out significant pair production if
a plasma process is responsible for coherence and
the radio emission originates from inside the light cylinder.

\end{abstract}
\twocolumn
\setcounter{footnote}{0}

\section{Introduction}
Pulsars are generally believed to be fast rotating neutron stars with
very strong magnetic fields. As shown by Goldreich and Julian (1969)
the rotating magnetic dipole produces a quadrupole electric field
whose component parallel to the open magnetic field lines at the poles
extracts particles very effectively from the neutron star surface and
accelerates them to highly relativistic energies. Thus the
magnetosphere is filled with an electron plasma which shields the
electric field. Complete shielding is established when the net charge
density reaches the value $\rho_{\rm GJ}=2B\Omega\varepsilon_0$ (the
so-called {\em Goldreich-Julian density}). For a single charge
magnetosphere this implies a particle density of $n_{\rm GJ}=\rho_{\rm
GJ}/e$. Hence the use of the  vacuum electric field is not realistic
for a model of pulsar particle acceleration but the residual potential
drop will have to be considered.
 If the Lorentz factors of the accelerated particles reach
about $10^6$ they emit hard curvature radiation that (after some
distance) propagates at a sufficient angle to the magnetic field so
that significant pair production can occur (Erber 1966). In this case
it is commonly accepted that the newly created particles produce more
pairs by emitting energetic synchrotron or curvature radiation. As a
result an avalanche of secondary particles populates the
magnetosphere with densities of about $10^4n_{\rm GJ}$
(e.g. Ruderman and Sutherland 1975; Daugherty  and Harding 1983;
Cheng and Ruderman 1977; Chang 1995).

The most detailed information about on pulsar magnetospheres
is obtained from radio observations. Pulsar radio emission is known to
be of highly coherent nature as the brightness temperatures reach up
to $10^{25}$ K and more (Sturrock 1971; Michel 1982, Lyne and
Graham-Smith 1990 and references therein) and it has been deduced
from the observations that the emission originates mostly well inside
the light cylinder (Blaskiewicz et al. 1991; Kijak and Gil 1997; Kramer
et al. 1997, von Hoensbroech et al. 1998a).
The location of the radio emission region deep in the pulsar
magnetosphere has significant consequences for the properties of the
magnetospheric plasmas with respect to the propagation of
electromagnetic waves. It is the aim of this contribution to show that
emission zones well inside the light cylinder imply that almost no
pair production in the inner region of pulsar magnetospheres is
allowed at least for the Crab pulsar. In other words, a too high
density population of secondary particles in a neutron star
magnetospheres is in contradiction to the observed low frequency
emission from such objects. 

\section{Limitations for the plasma density} 
Our central argument relies on the fact that pulsar radio emission
is of coherent nature.
Any coherent process require
phase coupling of the radiating particles. The most natural way for a
plasma in a strong magnetic field to force particles to emit
coherently by longitudinal plasma fluctuations. Such plasma waves are
created by density fluctuations because the charges are displaced
along the field lines in order to cancel the electric fields caused by
the fluctuations. The characteristic frequency of these waves just
depends on the density of the plasma and its Lorentz factor and is
called {\em plasma frequency}: 
\begin{equation}
\nu_{\rm pe}=\frac{1}{2\pi}
\sqrt{\frac{ne^2}{m_{\rm e}\varepsilon_0\gamma}}\,.
\end{equation}
This frequency is frame invariant (Ruderman and Sutherland 1975).
If we now assume a plasma process (like a two-stream instability)
being the cause for bunching  (which is necessary to get the phase
coupling and therefore coherence) then it is required that the emitted
waves have a frequency of at least the local value of $\nu_{\rm pe}$
in the beam frame
\footnote{This is strictly the case for any coherence mechanism that
involves Langmuir waves. Although magnetoacoustic waves can  propagate
at frequencies below the $\nu_{\rm pe}$ (Volokitin et al. 1985,  Arons
\& Barnard, 1986) they are an unlikely cause of the bulk of  coherent
radio emission. It is not clear how the prodigious observable radio
fluxes can be generated in the form of these waves. Furthermore their
polarisation properties expect us to find predominantly strongly
linearly polarised sources if one excludes the possibility of a
strongly depolarising medium between the point where magnetoacoustic
wave transforms into e.m. radiation and the region where the cycloton
resonance occurs.}.
In our general discussion we do not yet fix a mechanism but just
assume that $\nu_{\rm obs}=\gamma^\alpha\nu_{\rm pe}$ with $\nu_{\rm
obs}$ being the observed frequency and $\alpha > 1$. Then the {\em
minimum observed} frequency reads 
\begin{equation}
\label{minobs}
\nu_{\rm min}=\nu_{\rm pe}\gamma^\alpha=\frac{1}{2\pi}
\sqrt{\frac{ne^2\gamma^{2\alpha-1}}{m_{\rm e}\varepsilon_{0}}}\,.
\end{equation}
Rewriting the density in units of the Goldreich-Julian density
$n_{\rm GJ}$ as 
\begin{equation}
  \xi=\frac{n}{n_{\rm GJ}} 
\end{equation}
eqn. (\ref{minobs}) becomes
\begin{equation}
\nu_{\rm min}=\frac{1}{2\pi}\sqrt{\frac{2B\Omega e}{m_{\rm e}x^3}}\,
\xi^{1/2}\gamma^{(2\alpha-1)/2}
\end{equation}
with $x=r/r_{\rm NS}$ denoting the distance of the emission region to
the centre of the star in pulsar  radii (hereafter called {\em
emission height}). This equation can be solved to obtain an expression
just containing observational data, fundamental constants and the two
interesting quantities $\xi$ and $\gamma$:
\begin{equation}
\xi\gamma^{2\alpha-1}=\frac{4\pi^2\nu_{\rm min}^2m_{\rm e}}{2B\Omega
e}x^3\,.
\end{equation}
Next we use the condition that the radio emission must take place
well inside the light cylinder $R_{\rm LC}=c/\Omega$. For the
most conservative estimate we use $x=x_{\rm LC}$ because this
is the maximum possible emission height.
The minimum frequency is estimated to be about equal to the frequency
where the highest radio flux is observed because most pulsar radio
spectra show their maxima near the lowest frequencies observable. This
frequency will be denoted by $\nu_{\rm Smax}$. Thus we
find an upper limit for $\xi\gamma^{2\alpha-1}$ by setting  $\nu_{\rm
min}=\nu_{\rm Smax}$. Inserting the above expressions we find
\begin{equation}
\label{maxdens}
(\xi\gamma^{2\alpha-1})_{\rm max}=\frac{2\pi^2\nu_{\rm Smax}^2m_{\rm
e}c^3}{B\Omega^4e\,r_{\rm NS}^3}.
\end{equation}
This value cannot be exceeded if an ordinary plasma process,  which
involves the plasma frequency, is responsible for the
coherence. In particular it provides an upper limit for the particle
density since the Lorentz factor always exceeds unity.

Our reasoning does not depend on the actual  possibility of
propagation of the generated radio waves. We consider only  the
consequences of coherence produced by longitudinal plasma fluctuations
at the plasma frequency, under the proviso that  the emission heights
are located inside the light cylinder, as deduced from common
observational evidence. 

\section{Discussion}
We apply the calculations above to the most prominent
pulsar observed in all frequency bands, PSR 0531+21 (also known as the
Crab pulsar). This neutron star has a rotational period of
$P=33.4\,$ms,  a surface magnetic field strength of
$B=3.8\cdot10^8\,$T and its radio emission has a maximum flux at
$\nu_{\rm Smax}=160\,$MHz. These values can now be inserted in
eqn. (\ref{maxdens}) to find the upper limit
\begin{equation}
  \xi\gamma^{2\alpha-1}=163.
\end{equation}
For {\bf any} plasma process that allows low frequency
emission the product of density and Lorentz factor cannot exceed the
value given above (since $\alpha\geq1$ as mentioned).
The condition that 160 MHz waves are produced by a plasma process
(all known highly coherent processes are of this type) in the Crab
pulsar requires either low densities (comparable to $n_{\rm GJ}$) and
therefore little pair production or very low Lorentz factors or
both. In consequence it is very likely that in {\em inner gaps} the
Lorentz factors cannot be too high since otherwise pair production
would set in.

Probably the situation is even worse than that: As shown by Lesch et
al. (1998) the idea that coherent curvature radiation
(for which $\alpha=1$) produces the radio emission must be rejected
since the energetics of this radiation mechanism does not reproduce
the observed pulsar luminosities at frequencies smaller than several
GHz. If we adopt the other currently favoured theory that inverse
Compton scattering with plasma solitons (free electron maser) is the
process causing radio emission (e.g. Melrose 1992) the constant
calculated above gives the maximum value for $\xi\gamma^3$, since for
that mechanism $\alpha=2$. Thus, even for low-relativistic
particles we could not allow for any significant pair production,
which is in accordance with the pair production models (e.g. Chang 
1995).

For "standard pulsars" (i.e. $\Omega\approx 4\pi/{\rm s}$, $B\approx
10^8\,{\rm T}$) the limit for the density would certainly be far
higher if we adopt the estimation $x_{\rm em}\approx x_{\rm LC}$. But
observations suggest the emission heights to be far inside the light
cylinder (Blaskiewicz et al. 1991; Kijak and Gil 1997; Kramer {\em
et al.} 1997; von Hoensbroech et al. 1998a) at about
$x_{\rm em}\approx 50$ whereas $x_{\rm LC}\approx 2000$. If we insert
this we obtain a limit of about $\xi\gamma^{2\alpha-1}\approx
5000$. The argument given above is applicable here too, although it is
not as strong as for Crab. 

Note that our arguments just concern the lowest radio frequencies.
Since all observations suggest that the radio emission to be
broad-band and highly localized in space and time the most likely
scenario is that the density does not change significantly but the
emitted frequency is fixed only by the Lorentz factor.
Our view is in accordance with von Hoensbroech et al. (1998b). They
showed that the complex overall features of polarization can be well
explained by a cold single charge model.
Finally we would like to note that our argumentation does not apply
to processes outside the radio emission region. In {\em outer
gaps} pair creation and highly energetic emission may be
possible. Therefore outer gaps or centrifugal forces may be
responsible for high energy radiation (X-rays, $\gamma$-rays) or the
highly relativistic particles needed for the energy supply of the
nebula (Romani, 1996). 

Finally we summarize our ideas and their implications.
Given the one-dimensionality of the particle motion along the
strong magnetic field, the only reliable mechanisms for maintaining
the necessary phase coupling and
coherence involve longitudinal plasma fluctuations, whose
characteristic frequency is the plasma frequency. 
This has profound implications for the particle number density and
energy in that sense that the product of both parameters should not
exceed a critical value. For the Crab pulsar, which has its maximal
radio flux at about 160 MHz this means that no population of
electron-positron pairs highly exceeding the Goldreich-Julian density 
can be present in the radio emission zone. 
Additionally the radiating particles must have low energies. Their
Lorentz factors can not much exceed 100. Since many pulsars exhibit
radio maxima at some 100 MHz our conclusion holds for a significant
part of the pulsar population. 

{\it Acknowledgements:} We are grateful to the referee Don Melrose for
drawing our attention to the possibility of magneto-acoustic wave
propagation below $\nu_{\rm pe}$.


\begin{references}{}

  \reference{} {Arons, J., Barnard, J.J., ApJ 302, 120, 1986}
  \reference{} {Blaskiewicz, M., Cordes, J.M., Wasserman, I., 1991,
               ApJ 370, 643}
  \reference{} {Chang, H.K.,1995, A\&A 301, 456}
  \reference{} {Cheng, A.F., Ruderman, M.A., 1977,  ApJ 214, 598}
  \reference{} {Daugherty, J.K., Harding, A.K., 1982, ApJ 252, 337}
  \reference{} {Erber, T, 1966 Rev. Mod. Phys. 38, 626}
  \reference{} {Goldreich, T, Julian, W.H., 1969, ApJ 157, 869}
  \reference{} {von Hoensbroech, A., Kijak, J., Krawczyk, A. 1998a,
               A\&A 334, 571}
  \reference{} {von Hoensbroech, A., Lesch, H., Kunzl, T. 1998b,
               A\&A 336, 209}
  \reference{} {Kijak, J., Gil, J. 1997, MNRAS 288(3), 631}
  \reference{} {Kramer, M. ,Xilouris, K.M., Jessner, A., Lorimer,
               D.R., Wielebinski, R. Lyne, A.G., 1997, A\&A 322, 846}
  \reference{} {Lesch, H., Jessner, A., Kramer, M. Kunzl, T., 1998,
               A\&A 332, L21}
  \reference{} {Lyne, A.G., Graham-Smith, F., 1990, Pulsar Astronomy,
               Cambridge University Press}
  \reference{} {Melrose, D.B., 1992, in: The Magnetospheric
               Structure and Emission Mechanisms of Radio Pulsars.
               eds. Hankins, T.H., Rankin, J.M. and Gil, J.A., IAU
               Colloq. {\bf 128}, 133}
  \reference{} {Michel, F.C., 1982, Rev. Mod. Phys. 54, 1}
  \reference{} {Romani, R., 1996, ApJ 470, 469}
  \reference{} {Ruderman, M.A., Sutherland, P.G., 1975, ApJ 196, 51}
  \reference{} {Sturrock, P.A., 1971, ApJ 164, 529}
  \reference{} {Volokitin, A.S., Krasnoselskikh, V.V., Machabeli,
		G.Z., Sov.J.Plasma Phys 11, 310, 1985};
 

\end{references}
\end{document}